\documentclass[aps,pre,twocolumn,superscriptaddress,10pt]{revtex4-2}
\pdfoutput=1

\usepackage{amsmath}
\usepackage{bm}
\usepackage{caption}
\usepackage{graphicx}
\usepackage{hyperref}
\usepackage{microtype}

\newcommand{\abs}[1]{|#1|}
\newcommand{\avg}[1]{\langle #1 \rangle}
\newcommand{\avgst}[1]{\langle #1 \rangle\st}
\renewcommand{\eqref}[1]{Eq.~(\ref{#1})}
\newcommand{\figref}[1]{Fig.~\ref{#1}}
\newcommand{\hg}[2]{{}_{#1}F_{#2}}
\newcommand{\st}{_\mathrm{st}}

\begin{document}

\title{Will a time-varying complex system be stable?}

\newcommand{\DFA}{Department of Physics and Astronomy ``Galileo Galilei'', University of Padova, Italy}
\newcommand{\INFN}{INFN, Padova Division, Italy}
\newcommand{\NBFC}{National Biodiversity Future Center, Palermo, Italy}

\author{Francesco Ferraro}
\email{francesco.ferraro.4@phd.unipd.it}
\affiliation{\DFA}
\affiliation{\INFN}
\affiliation{\NBFC}

\author{Christian Grilletta}
\affiliation{\DFA}

\author{Amos Maritan}
\affiliation{\DFA}
\affiliation{\INFN}
\affiliation{\NBFC}

\author{Samir Suweis}
\affiliation{\DFA}
\affiliation{\INFN}

\author{Sandro Azaele}
\affiliation{\DFA}
\affiliation{\INFN}
\affiliation{\NBFC}

\begin{abstract}
Randomly-assembled dynamical systems are theoretically predicted to be unstable upon crossing a critical threshold of complexity, as first shown by May. Yet, empirical complex systems exhibit remarkable stability, indicating the presence of additional mechanisms playing a stabilizing role. The relation between complexity and stability is typically assessed by assuming fixed interactions, whereas real systems often evolve in intrinsically time-dependent states. To understand how this affects stability, we linearize a general non-autonomous dynamics around a reference operating state and model the resulting parameters as stochastic processes, which represent the minimal extension of static random interactions to time-varying ones. We derive exact stability bounds that generalize complexity-stability theory to dynamically varying systems. Notably, we find that temporal variability allows systems to remain stable even when their instantaneous Jacobian would predict instability. We compare our results against a non-linear neural network model, where our theory applies exactly, and the generalized Lotka-Volterra equations, where we numerically find that time-varying interactions systematically postpone the onset of replica-symmetry breaking. Overall, our results indicate that temporal variability systematically improves stability, demonstrating a general mechanism by which complex systems can violate classical complexity-stability bounds.
\end{abstract}

\maketitle

\section{Introduction}

Na\"ively, one might expect that a system with more interacting components, or with more diverse and numerous connections, should be more stable. This intuition was challenged by May in his seminal work ``Will a large complex system be stable?'' \cite{may1972will}, which established that complexity can instead beget instability \cite{fyodorov2016nonlinear}. The apparent contradiction between this result and the striking stability of empirical complex systems \cite{albert2000error,ives2007stability,battiston2016complexity,cimini2019statistical} prompted the so-called complexity-stability debate \cite{mccann2000diversity,allesina2012stability,landi2018complexity,jacquet2016no,yonatan2022complexity}. While May-like arguments rely on linear analysis, recent works have shown that these qualitative conclusions hold in more general settings. Specifically, in non-linear models with random interactions, increasing complexity generically drives systems toward instability or marginal stability, as demonstrated in both ecological \cite{bunin2017ecological,barbier2018generic,biroli2018marginally,stone2018feasibility,gibbs2018effect,altieri2021properties,hu2022emergent,ros2023generalized} and neural \cite{sompolinsky1988chaos,kadmon2015transition,schuecker2018optimal,martorell2024dynamically,albanese2024replica} models.

A sizable body of literature, especially in theoretical ecology, has sought to identify which additional stabilizing structures could reconcile theory with observation \cite{haldane2011systemic,suweis2014disentangling,tu2019reconciling,pettersson2020stability,chen2024stability}. Complexity-stability bounds are usually obtained via local stability analysis around fixed states, with structural constraints imposed on the static coefficients of the resulting linear dynamics to assess whether they enhance or hinder stability relative to a completely random null model.

In many empirical systems, however, interactions are intrinsically time-dependent, making the resulting dynamics non-autonomous. For example, the plasticity of synaptic connections continuously modulates operating conditions of nervous systems, which underlies memory and learning \cite{sjostrom2001rate,magee2020synaptic,clark2024theory}. In ecological communities, dynamically fluctuating species interactions are a key factor driving ecosystem properties \cite{fiegna2015evolution,ushio2018fluctuating,pacciani2020dynamic}. Despite the ubiquity of such time-dependent interactions \cite{scire2005dynamic,goh2005nonlocal,zhou2006dynamical,hu2012blood,lonardi2023infrastructure,caligiuri2023lyapunov,caligiuri2025characterizing}, and general results indicating that temporal variability can alter system stability relative to what the instantaneous Jacobian predicts \cite{chen1984linear,rugh1996linear,khalil2002nonlinear}, a systematic understanding of how time-varying interactions affect the stability of complex systems remains lacking \cite{lotka1921note,botkin1975stability,cohen1984stability,kondoh2003foraging,dakos2019ecosystem,tian2020estimating,krumbeck2021fluctuation,al2026spatiotemporal,henderson2026fluctuating}.

In this work, we address this gap by assessing the stability of a complex system operating in a time-dependent state. We consider a generic non-autonomous dynamics and linearize it around a reference operating state. The resulting parameters are modeled as colored noises, the natural generalization of static random variables to temporally fluctuating ones. Building on recent advances in disordered systems with stochastic interactions \cite{suweis2024generalized,ferraro2025exact,zanchetta2025emergence,gupta2025non}, we analytically derive complexity bounds which show that temporal variability systematically enhances complex dynamics. Remarkably, we find that a system can remain dynamically stable beyond the May bound, where its instantaneous Jacobian has eigenvalues with positive real part.

We support our analytical results with numerical simulations of two paradigmatic complex systems: a neural network model \cite{sompolinsky1988chaos}, where our linear framework applies exactly, and the generalized Lotka-Volterra equations \cite{hofbauer1998evolutionary,bunin2017ecological}, a foundational model in theoretical ecology. For the latter, we numerically observe the same phenomenology as the linear model, that is, temporal variability extends the regime of stability against replica-symmetry breaking.

Our results point to a general principle: time-varying interactions can lead to more stable complex systems than their static counterparts. This may ease the long-standing tension between theoretical and empirical stability.

\section{Setup}

We consider a large system following a generic first-order non-autonomous dynamics
\begin{equation}
    \dot{x}_i(t) = f_i(\bm{x}(t),t),
    \label{eq:ODE}
\end{equation}
where $\bm{x}(t) =(x_1(t),\dots,x_N(t))$ is the state vector of the system, which is composed of $N\gg1$ degrees of freedom. We introduce the reference state $\bm{x}^*(t)$, defined implicitly by $f_i(\bm{x}^*(t),t)=0$ for all $i, t$. Under mild regularity conditions, $\bm{x}^*(t)$ exists locally and varies smoothly in time  \footnote{Assuming sufficient regularity, differentiating $\bm{f}(\bm{x}^*(t),t)=\bm{0}$ yields 
$\dot{\bm{x}}^*(t)=-M^{-1}(t)\,\partial_t\bm{f}(\bm{x}^*(t),t)$, providing a closed equation for $\bm{x}(t)$, whenever $M(t)$ is invertible. Moreover, $\bm{h}(t)=-\dot{\bm{x}}^*(t)=M^{-1}(t)\partial_t\bm{f}(\bm{x}^*(t),t)$.}. Small perturbations around this state $\bm{\delta x}(t)=\bm{x}(t)-\bm{x}^*(t)$ evolve according to
\begin{equation}
    \delta \dot{x}_i(t) = \sum_j M_{ij}(t) \delta x_j(t) + h_i(t),
    \label{eq:model}
\end{equation}
where
\begin{equation}
    M_{ij}(t) = \frac{\partial f_i}{\partial x_j} (\bm{x}^*(t),t)
\end{equation}
is the instantaneous Jacobian matrix evaluated at the reference state and $h_i(t) = - \dot{x}^*_i(t)$. \eqref{eq:model} admits a straightforward interpretation. The interaction term is the instantaneous linear response around the reference state, generalizing the autonomous case. The term $h_i(t)$ is instead a comoving-frame contribution: as the reference state $\bm{x}^*(t)$ evolves, the origin of the linearized coordinates shifts in time, which acts as an effective forcing.

A fundamental complexity-stability bound was first derived by May \cite{may1972will}. He considered an autonomous linearized dynamics, with constant coefficients $M_{ij}$ and no forcing, $h_i=0$. He assumed $M_{ii}=-1$ for all $i$ and that the off-diagonal entries of the Jacobian are drawn independently from a random distribution with zero mean. Random matrix theory, in particular the circular law \cite{ginibre1965statistical,girko1985circular,sommers1988spectrum,tao2010random,livan2018introduction}, then implies that the spectrum of the Jacobian matrix in the complex plane is uniformly supported on a disk centered at $(-1,0)$ with radius $\sigma\sqrt{NC}$, where $\sigma$ is the standard deviation of the off-diagonal distribution and $C$ is the connectance (fraction of non-zero entries). From this, May concluded that an equilibrium becomes linearly unstable whenever $\sigma\sqrt{NC}>1$. 

May's caricature of a complex system serves as a null model that can be extended to determine which features promote or suppress stability \cite{allesina2012stability,chen2024stability}. In this spirit, we construct a minimal dynamical extension by promoting the random variables $M_{ij}$ and $h_i$ to time-dependent stochastic processes. Specifically, we set $M_{ii}(t)=-1$ for all $i$, $t$, and model the off-diagonal elements $M_{ij}(t)$ and $h_i(t)$ as independent Gaussian processes with zero mean and correlations
\begin{equation}
\begin{aligned}
    \avg{M_{ij}(t) M_{ij}(t')} &= \frac{\sigma^2}{N} Q(t-t'), \\
    \avg{h_i(t) h_i(t')} &= \rho^2 Q(t-t'), \\
    Q(t) &=e^{-\abs{t}/\tau},
\end{aligned}
\label{eq:model-correlations}
\end{equation}
where $\sigma$ and $\rho$ are fixed parameters, and $\tau$ is the correlation time. The mean of $h_i(t)$ can be set to zero without loss of generality, as a non-zero mean does not affect the stability threshold derived below. The specific correlation profile set by $Q(t)$ yields Ornstein-Uhlenbeck processes, which represent the unique processes that are both Gaussian and Markovian \cite{doob1942brownian}. Such time-varying interactions are sometimes referred to as ``annealed'' \cite{jain2017diffusing,khadem2021transport,santra2022effect,suweis2024generalized,ferraro2025exact,zanchetta2025emergence,jkedrzejewski2025does}, in contrast to quenched interactions that are frozen in time \cite{mezard1987spin,azaele2024generalized,zenari2026generalized}. A fraction of the matrix entries is set to zero at all times, fixing the network connectance to $C<1$. 

Unlike in May's approach, the interactions in \eqref{eq:model-correlations} are scaled by $N$ to guarantee a well-behaved large-$N$ limit \cite{cohen1984stability}. To map our results back to May's framework and related literature, one can substitute $\sigma\sqrt{C}\to\sigma\sqrt{NC}$ in the results that follow.

For simplicity of presentation, we assume that the interactions and forcing terms have the same correlation time $\tau$. The general case is detailed in SM, where we show that stability is governed solely by the fluctuations of $M_{ij}(t)$. Although in a fully specified non-autonomous system $\bm{h}(t)$ may be correlated with $M(t)$, we neglect such correlations. Since the statistics of the forcing terms will not enter the stability conditions, this assumption is not expected to qualitatively change our conclusions. Furthermore, while we assume independent entries $M_{ij}(t)$, numerical results presented in the SM address a more general case with reciprocally correlated entries and confirm the phenomenology presented here.

The model of May is recovered in the limit $\tau\to\infty$, up to a rescaling of the interactions. At finite correlation time, the resulting dynamical system differs from that of May. Still, at any fixed time, the spectral radius of the Jacobian matrix $M_{ij}(t)$ remains $R=\sigma\sqrt{C}$ for any value of $\tau$ and $t$. In particular, for $R>1$ some eigenvalues of the instantaneous Jacobian matrix have positive real part.

\section{Stability criterion}

The linear model defined by \eqref{eq:model} and \eqref{eq:model-correlations} either converges to a stable stationary state or displays diverging solutions. In the former case, fluctuations around the reference state remain bounded, indicating a stable system, while in the latter the system is linearly unstable. Rather than solving \eqref{eq:model} directly, we consider the associated Dynamical Mean-Field Theory (DMFT) equation \cite{sompolinsky1982relaxational,binder1986spin,mezard1987spin}, which describes the dynamics of a representative degree of freedom $\delta x(t)$ of the system in the large-$N$ limit.  

The DMFT equation corresponding to \eqref{eq:model} reads
\begin{equation}
    \delta \dot{x}(t) = -\delta x(t) + R\eta(t) + h(t),
    \label{eq:DMFT}
\end{equation}
where $R=\sigma\sqrt{C}$, and $\eta(t)$, $h(t)$ are independent Gaussian noises with zero mean and correlations
\begin{equation}
\begin{aligned}
    \avg{\eta(t)\eta(t')} &= Q(t-t') \avg{\delta x(t)\delta x(t')}, \\
    \avg{h(t)h(t')} &= \rho^2 Q(t-t').
\end{aligned}
\end{equation}
These averages are understood to be over all realizations of $\eta(t)$ and $h(t)$. The derivation of \eqref{eq:DMFT} follows closely the procedure detailed in \cite{suweis2024generalized,ferraro2025exact,zanchetta2025emergence} and is omitted here. We note that the DMFT \eqref{eq:DMFT} implies that stability depends on the same complexity parameter $R=\sigma\sqrt{C}$ originally identified by May (up to a rescaling of the interactions), together with the additional parameters $\rho$ and $\tau$, the latter entering via $Q(t)$. 

By multiplying the DMFT~\eqref{eq:DMFT} by $h(0)$, taking the stationary average, and using the fact that $\eta(t)$ and $h(t)$ are independent, we obtain a closed equation for the stationary cross-correlation $D(t)=\avgst{\delta x(t)h(0)}$, which reads $\dot{D}(t) = -D(t) + \rho^2 Q(t)$. We then rewrite \eqref{eq:DMFT} as $R\eta(t) = \delta \dot{x}(t) + \delta x(t) - h(t)$ and average the product $R^2 \eta(t)\eta(t')$. After performing straightforward manipulations, taking the stationary limit, and employing the equation for $D(t)$, we find a closed equation for the stationary autocorrelation $C\st(t) = \avgst{\delta x(t) \delta x(0)}$:
\begin{equation}
    -\ddot{C}\st(t) + \big[1 - R^2 Q(t)\big]\,C\st(t) = \rho^2 Q(t).
    \label{eq:correlation-ODE}
\end{equation}

\eqref{eq:correlation-ODE} is supplemented with the boundary conditions $\dot{C}\st(0)=0$, by time-reversal symmetry, and finite $C\st(\infty)$, assuming a bounded stationary state. The solution can be obtained by standard methods (see SM). Evaluating $C\st(t)$ at $t=0$ yields the stationary variance of \eqref{eq:model}:
\begin{equation}
    \sigma\st^2 = 
    \frac{\rho^2}{R^2}
    \left[
    \frac
        {\hg{1}{2}(\tau;1+\tau,1+2\tau;-\tau^2R^2)}
        {2\hg{0}{1}(2\tau;-\tau^2R^2)-\hg{0}{1}(1+2\tau;-\tau^2R^2)}
    -1
    \right],
    \label{eq:variance}
\end{equation}
where $\hg{p}{q}$ is the generalized hypergeometric function \cite{weisstein2006generalized}. It also turns out that $C\st(\infty)=0$ (see SM).

The divergence of the stationary variance marks the onset of instability in the system, which takes place when the denominator in \eqref{eq:variance} first vanishes. This condition yields the critical complexity $R$ of the model above which the linear system displays diverging solutions. Using Eq.~(9.1.69)~of~\cite{abramowitz1972handbook}, this condition can be rewritten as $R J_{2\tau-1}(2\tau R) - J_{2\tau}(2\tau R) = 0$, where $J_n(x)$ is the Bessel function. Using then Eq.~(9.1.27)~of~\cite{abramowitz1972handbook}, this condition can be further expressed as the smallest positive $R$ satisfying
\begin{equation}
    J'_{2\tau}(2\tau R) = 0.
    \label{eq:stability-bound}
\end{equation}
We note that stability is independent of the parameter $\rho$. Furthermore, in the general case of different correlation timescales for $M_{ij}(t)$ and $h_i(t)$, stability is determined solely by the correlation time of the interaction matrix (see SM).

\begin{figure}
    \centering
    \includegraphics[width=\linewidth]{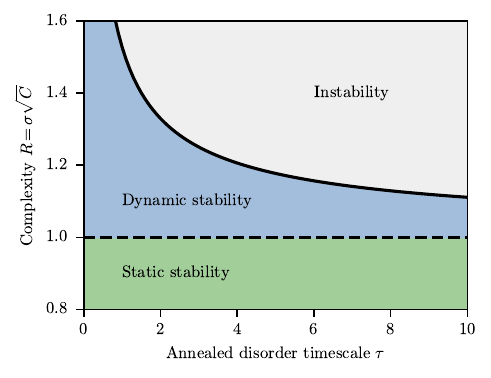}
    \caption{Phase diagram of the linear model \eqref{eq:model} with annealed interactions as set by \eqref{eq:model-correlations}. Below the bound given by \eqref{eq:stability-bound} (solid line), the model converges to a stable stationary state, while above it the system is unstable. The two stable phases differ in that, below the May bound $R=1$ (dashed line) in the ``Static stability'' region, all eigenvalues of the instantaneous Jacobian have negative real part. Instead, above the May bound, some eigenvalues of the Jacobian have positive real part, yet the dynamics is stabilized by sufficiently rapid temporal variability, in the ``Dynamic stability'' phase. The critical complexity below which the system remains dynamically stable increases as the scale of temporal variability $\tau$ is reduced, and diverges as $\tau\to0$.}
    \label{fig:phase-diagram}
\end{figure}

The solution of \eqref{eq:stability-bound} is shown in \figref{fig:phase-diagram}. The critical complexity required for stability is always above the May bound $R=1$, as it follows from the fact that the smallest zero of $J'_n(x)$ is always larger than $n$ (see Section~15.3 of \cite{watson1944treatise}). This demonstrates that temporal variability systematically promotes stability. In the ``Dynamic stability'' region, moreover, the system is stable even though a fraction of the eigenvalues of the instantaneous Jacobian have positive real part. The critical complexity also increases as the timescale of temporal variability $\tau$ is reduced, indicating that faster temporal fluctuations further enhance stability. In SM we provide a qualitative explanation of how temporal variability in the interactions can stabilize a fixed point.

\section{Non-linear models}

We illustrate our results in simulations of non-linear models. We examine two different settings: a neural network model, where the linear theory applies exactly, and the GLV equations, where temporal fluctuations suppress the onset of replica-symmetry breaking \cite{mezard1987spin,bunin2017ecological,biroli2018marginally,altieri2021properties}.

\begin{figure}
    \centering
    \includegraphics[width=\linewidth]{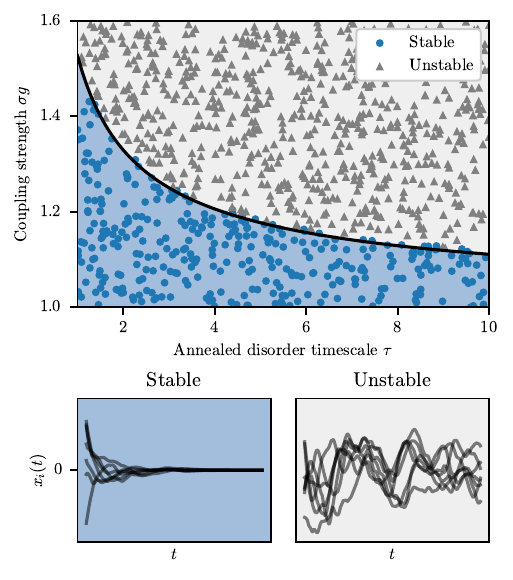}
    \caption{Validation of the stability bound in a neural network model with annealed interactions. In the upper panel we compare the critical complexity threshold \eqref{eq:stability-bound} with $R=\sigma g$ (solid line) with numerical integration of the model \eqref{eq:NN} with interactions given by \eqref{eq:NN-interactions} (circles and triangles). Each realization is classified as stable if it settles into the fixed point $x_i=0$, and unstable if it reaches a non-zero stationary state. The agreement is excellent, as expected since our stability bound applies exactly to this model. The lower panels qualitatively display the dynamics in the two phases.}
    \label{fig:neural-network}
\end{figure}

First, we consider a firing-rate model of a network of $N$ neurons. We assume the synaptic current $x_i(t)$ evolves according to the non-linear dynamics \cite{sompolinsky1988chaos,kadmon2015transition,schuecker2018optimal}
\begin{equation}
    \dot{x}_i(t) = -x_i(t) + \sum_{j \neq i} J_{ij}(t) \phi(x_j(t)),
    \label{eq:NN}
\end{equation}
where the transfer function is, for definiteness, $\phi(x)=\tanh(gx)$. We assume an annealed interaction matrix, with full connectivity $C=1$ and such that
\begin{equation}
    \avg{J_{ij}(t) J_{ij}(t')} = \frac{\sigma^2}{N} Q(t-t').
    \label{eq:NN-interactions}
\end{equation}
 At small heterogeneity $\sigma$, the stationary state $x_i(t)=0$ is stable. Perturbations around this state follow the linear dynamics
\begin{equation}
    \delta \dot{x}_i(t) = -\delta x_i(t) + \sum_{j \neq i} J_{ij}(t)g \delta x_j(t).
\end{equation}
This corresponds exactly to the linear system studied previously via the identification $R=\sigma g$ and $h_i(t)=0$. The latter condition reflects a time-invariant stationary state. The stability condition is then given by \eqref{eq:stability-bound}. Numerical simulations confirm the agreement with the theoretical prediction, see \figref{fig:neural-network}.

Next, we consider an ecological community governed by the GLV equations \cite{hofbauer1998evolutionary,bunin2017ecological}. The abundance $x_i(t)$ of each species evolves according to
\begin{equation}
    \dot{x}_i(t) = x_i(t) \left[1 - x_i(t) + \sum_{j \neq i} \alpha_{ij}(t) J(x_j(t))\right].
    \label{eq:GLV}
\end{equation}
To avoid unbounded growth, we introduced a saturating response function $J(x)$ \cite{sidhom2020ecological,suweis2024generalized,zenari2026generalized}. The interaction matrix is annealed, and normalized as \cite{suweis2024generalized}
\begin{equation}
    \alpha_{ij}(t) = \frac{\mu}{N} + \frac{\sigma}{\sqrt{N}} z_{ij}(t),
\end{equation}
where $\avg{z_{ij}(t) z_{ij}(t')} = Q(t-t')$. For this model, our previously derived linear bound is not expected to hold, as the Jacobian matrix around the instantaneous reference state is not described simply by \eqref{eq:model-correlations}, and moreover it is correlated with the external field $h_i(t)$.

Unlike the neural model, this system possesses no time-invariant fixed point. We therefore assess the stability of the community by considering two replicas of the community $\bm{x}^a$ and $\bm{x}^b$. The two replicas are subjected to the same realization of the disorder $\alpha_{ij}(t)$ but have different initial conditions. We monitor the distance between replicas, defined as
\begin{equation}
    d(t) = \frac{1}{N} \sum_i \left[x^a_i(t) - x^b_i(t)\right]^2.
    \label{eq:d(t)}
\end{equation}
If $d(t)$ converges to zero, the system displays replica-symmetry (RS) and reaches a stationary state stable to perturbations. Conversely, if $d(t)$ remains non-zero, replica-symmetry is broken (RSB). These two phases are the annealed counterparts of the Unique Fixed Point and Multiple Attractor phases found in the quenched, $\tau\to\infty$ limit \cite{bunin2017ecological,sidhom2020ecological,zenari2026generalized}.

\begin{figure}
    \centering
    \includegraphics[width=\linewidth]{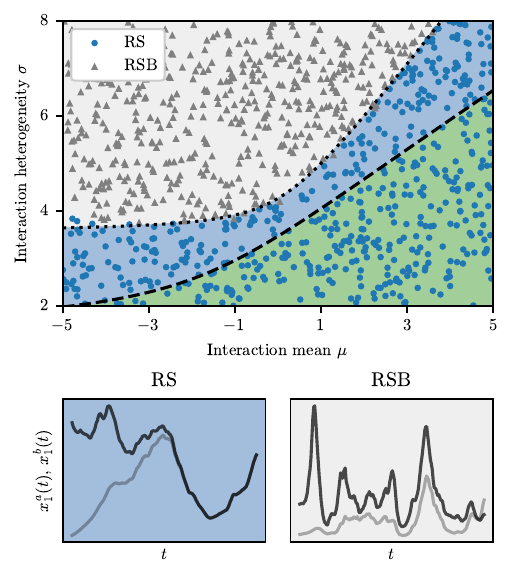}
    \caption{Numerical phase diagram of generalized Lotka-Volterra equations with annealed interactions. For each parameter set, we simulate two replicas of the GLV dynamics \eqref{eq:GLV} which share the same disorder realization $\alpha_{ij}(t)$ but start from different initial conditions. The system is classified as replica-symmetric (RS) if the distance $d(t)$ [\eqref{eq:d(t)}] vanishes at long times, and replica-symmetry-breaking (RSB) otherwise. The dashed line marks the stability boundary for the quenched case \cite{zenari2026generalized}, while the dotted line marks the numerical boundary separating the RS and RSB phases. As in \figref{fig:phase-diagram}, the lower green region indicates static stability, while the intermediate blue area indicates dynamic stability (see text). Consistent with the linear model, temporal variability extends the stable RS regime beyond the quenched limit. The response function is $J(x)=2x/(2+x)$ and the correlation time is $\tau=1$. The lower panels qualitatively show single species trajectories for the two replicas in each phase.}
    \label{fig:GLV}
\end{figure}

Results from the numerical integration of \eqref{eq:GLV} are shown in \figref{fig:GLV}. Both the lower (green) and intermediate (blue) regions are RS regimes, while the upper (gray) region is a RSB phase. The lower, dashed line marks the instability threshold in the quenched case. The difference between the two RS phases lies in the fact that, in the lower region, the system is additionally statically stable: if the interaction matrix were frozen, trajectories would relax to a stable fixed point, consistently with a May-like criterion based on the instantaneous Jacobian. In the blue region the system is only dynamically stable: the dynamics is RS under time-dependent interactions, yet the corresponding system after freezing the interaction matrix would not settle into a fixed point but display instead chaotic dynamics. Thus, temporal variability stabilizes the dynamics in a broader sense than fixed-point stability, extending beyond the regime where a direct May-style comparison applies.

Comparing these results with \figref{fig:phase-diagram} shows that the GLV phenomenology qualitatively mirrors the linear model: time-variability in interactions enhances stability relative to the quenched case by enlarging the parameter region in which replica symmetry is maintained. We note that this increased stability is not due to a reduction in the number of surviving species; indeed, as shown in \cite{suweis2024generalized}, annealed disorder actually increases the number of survivors, indicating that the stabilization mechanism is instead driven by temporal variability in the interactions.

\section{Discussion}

In this work, we constructed a minimal extension of the null model of May by promoting quenched random couplings to annealed stochastic processes. For this model we derived an exact complexity-stability criterion, which is given by \eqref{eq:stability-bound} and shown in \figref{fig:phase-diagram}. Two features are worth emphasizing. First, the stability region expands as the correlation time $\tau$ decreases, suggesting that faster temporal variability in the interactions promotes stability. Second, for any finite $\tau$ the stability boundary is strictly increased relative to the May one, recovered in the $\tau\to\infty$ limit.

Our results indicate that stability is controlled by the statistics and temporal correlations of the Jacobian $M_{ij}(t)$. The external forcings $h_i(t)$ act as an additive drive, setting the amplitude of fluctuations around the reference state, but do not shift the stability boundary (see also SM for generalizations). This implies that stabilization originates from temporal variability of the interactions rather than from details of the forcing.

We illustrated the mechanism in two paradigmatic non-linear settings. In the neural network model, the theoretical bound agrees with simulations, as expected since the linear description applies exactly. In the generalized Lotka-Volterra dynamics, time-dependent interactions postpone the transition to the replica-symmetry-breaking phase relative to the quenched case, indicating that the stabilizing effect of temporal variability extends beyond the exactly solvable linear model. An analytical characterization of the RSB phase for generic non-linear systems with time-varying interactions may be obtained by combining the present annealed-disorder framework with replicated DMFT approaches \cite{derrida1987dynamical,schuecker2018optimal,helias2020statistical}.

As in the original work of May, our model is intentionally stylized. Nevertheless, the stabilization mechanism admits a clear physical interpretation, see SM: rapid temporal fluctuations prevent growth along unstable directions by continuously reshuffling the instantaneous Jacobian eigenmodes. Empirical complex systems, which do not vary randomly but according to precise functional patterns, may harness this mechanism even more efficiently, potentially achieving stability levels well above the predictions of our null model. Notably, the stabilization we identified arises in the absence of any feedback of the system state onto interaction strengths.

Overall, temporal variability can stabilize complexity even when static criteria would predict instability, providing a generic route to overcome classical complexity-stability bounds. Future work should clarify how annealed disorder combines with the structural features explored in the complexity-stability debate \cite{allesina2012stability,landi2018complexity,chen2024stability}, and connect the theory to empirical settings by relating measurable interaction statistics and timescales to observed stability in neural \cite{sompolinsky1988chaos}, ecological \cite{bunin2017ecological}, and socio-economic systems \cite{haldane2011systemic,moran2019may}.

\section{Acknowledgments}
We thank Tommaso Jack Leonardi for insightful discussions.
FF acknowledges financial support under the National Recovery and Resilience Plan (NRRP), Mission 4, Component 2 Investment 1.4 - Call for tender No. 3138 of 16 December 2021, rectified by Decree n.3175 of 18 December 2021 of Italian Ministry of University and Research funded by the European Union - NextGenerationEU; Award Number: Project code CN00000033, Concession Decree No. 1034 of 17 June 2022 adopted by the Italian Ministry of University and Research, CUP C93C22002810006, Project title ``National Biodiversity Future Center - NBFC''. 
CG, AM and SA acknowledge financial support under the National Recovery and Resilience Plan (NRRP), Mission 4, Component 2, Investment 1.1, Call for tender No. 104 published on 2.2.2022 by the Italian Ministry of University and Research (MUR), funded by the European Union – NextGenerationEU – Project Title ``Emergent Dynamical Patterns of Disordered Systems with Applications to Natural Communities'' – CUP 2022WPHMXK - Grant Assignment Decree No. 2022WPHMXK adopted on 19/09/2023 by the Italian Ministry of Ministry of University and Research (MUR).
SS acknowledges financial support from the MUR-PNC (DD No. 1511 30-09-2022) Project No. PNC0000002, DigitAl lifelong pRevEntion (DARE)

\bibliography{refs}

\clearpage
\onecolumngrid
\appendix
\begin{center}
    \textbf{Supplemental Material} \\
    \phantom{.} \\
    \textbf{Will a time-varying complex system be stable?} \\
    Francesco Ferraro$^*$, Christian Grilletta, Amos Maritan, Samir Suweis, Sandro Azaele \\
    $^*$ \href{mailto:francesco.ferraro.4@phd.unipd.it}{francesco.ferraro.4@phd.unipd.it}
\end{center}

\section{Qualitative explanation for stabilization}

\begin{figure}[h]
    \centering
    \includegraphics[width=0.9\linewidth]{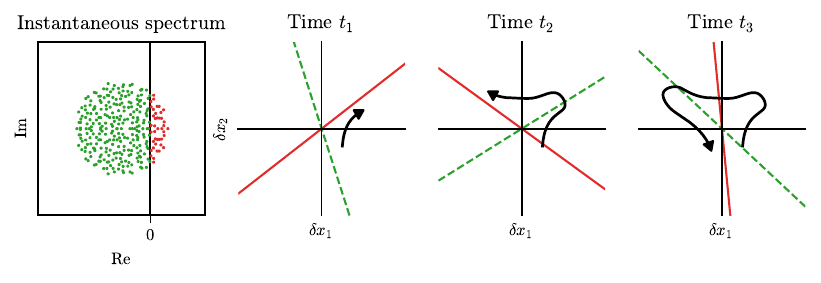}
    \caption{Qualitative explanation for the stabilization mechanism in time-varying systems. Left: The instantaneous spectrum of the Jacobian matrix $M_{ij}(t)$ is constant, and a fraction of this spectrum can have positive real part (red points). Right: Dynamics of perturbations at different times (curved arrow). The stable (green dashed lines) and unstable (red solid lines) directions continuously shift as the Jacobian matrix evolves, which effectively averages out the divergence. The panels are illustrative, as eigenvectors are generally complex.}
    \label{fig:qualitative}
\end{figure}

Let us briefly explain qualitatively how time-variability can stabilize a fixed point beyond static bounds, see \figref{fig:qualitative}. According to \eqref{eq:model-correlations} of the main text, the instantaneous spectrum of the Jacobian matrix $M_{ij}(t)$ is constant, and follows the circular law. When $R>1$ a fraction of its eigenvalues have positive real part. However, since the interaction matrix is time-dependent, the corresponding eigenvectors, and thus the directions of the unstable manifold, are continuously changing. Note that \figref{fig:qualitative} is strictly schematic, as eigenvectors are generally complex. As these unstable directions shift, the system never spends enough time in any single direction for a perturbation to grow unbounded. Ultimately, this averages out the divergence, allowing the system to remain dynamically stable even when the instantaneous Jacobian would predict static instability.

\section{Solution of \eqref{eq:correlation-ODE} of the main text}

We solve here the ODE \eqref{eq:correlation-ODE} of the main text. It is simpler to solve for the function $E(t)=C\st(t) + \rho^2/R^2$. This function satisfies the ODE
\begin{equation}
    -\ddot{E}(t) + \left[1 - R^2Q(t)\right]E(t) = \frac{\rho^2}{R^2}
    \label{SMeq:E}
\end{equation}
Consider the change of variables $z=2\tau R e^{-t/2\tau}$. This yields the equivalent homogeneous ODE
\begin{equation}
    z^2 E''(z) + z E'(z) + \left[z^2-(2\tau)^2\right]E(z)=0
\end{equation}
which is Bessel differential equation. Two independent solution of the homogeneous ODE of \eqref{SMeq:E} are then
\begin{equation}
\begin{aligned}
    u_1(t) &= J_{2\tau}(2\tau R e^{-t/2\tau}) \\
    u_2(t) &= Y_{2\tau}(2\tau R e^{-t/2\tau})
    \label{SMeq:u1-u2}
\end{aligned}
\end{equation}
Since the Wronskian of $u_1(t)$ and $u_2(t)$ is $1/\pi\tau$ then the general solution of \eqref{SMeq:E} is by the method of variation of parameters
\begin{equation}
    E(t) = c_1u_1(t) + c_2u_2(t) + \pi\tau \frac{\rho^2}{R^2} [u_2(t) U_1(t) - u_1(t) U_2(t)]
    \label{SMeq:E-solution}
\end{equation}
where $U_1(t)$ and $U_2(t)$ are the primitive functions of $u_1(t)$ and $u_2(t)$ respectively. Explicitly these are
\begin{equation}
\begin{aligned}
    U_1(t) &= - \frac{(\tau R)^{2\tau}}{\Gamma(1+2\tau)} e^{-t} \hg{1}{2}(\tau; 1+2\tau,1+\tau; - \tau^2R^2e^{-t/\tau}) \\
    U_2(t) &= \frac{(\tau R)^{2\tau} \cos(2\pi\tau) \Gamma(-2\tau)}{\pi} e^{-t} \hg{1}{2}(\tau; 1+2\tau,1+\tau; -\tau^2R^2 e^{-t/\tau}) \\
    &\qquad- \frac{(\tau R)^{-2\tau}\Gamma(2\tau)}{\pi} e^{t} \hg{1}{2}(-\tau; 1-2\tau,1-\tau; -\tau^2R^2 e^{-t/\tau})
\end{aligned}
\end{equation}
The primitive were found found be with the substitution $z = 2\tau R e^{-t/2\tau}$ and known integrals of Bessel functions. We want to impose the boundary conditions, which are that $\dot{E}(0)=0$ and that $E(\infty)$ is finite. To do this we use the asymptotic expansions of each term in \eqref{SMeq:E-solution} which are
\begin{equation}
\begin{aligned}
    u_1(t) &\sim \frac{(\tau R)^{2\tau}}{\Gamma(1+2\tau)} e^{- t} \\
    u_2(t) &\sim - \frac{\Gamma(2\tau)(\tau R)^{-2\tau}}{\pi} e^{t} \\
    u_1(t) U_2(t) &\sim - \frac{1}{2\pi\tau} \\
    u_2(t) U_1(t) &\sim \frac{1}{2\pi\tau}
\end{aligned}
\end{equation}
To have a finite $E(\infty)$ we impose that $c_2=0$. With this we also get that $E(\infty) = \rho^2/R^2$. From this we see that $C\st(\infty) = 0$, which indicates that the process $x(t)$ decorrelates with itself after a long time, which is expected. Imposing $\dot{E}(0)=0$ fixes the value of $c_1$, which is not reported here. This overall fixes the solution $E(t)$ and, consequently, $C\st(t)$. Setting $t=0$ in the solution found one obtains $E(0)$ and consequently  the stationary variance $C\st(0)$. After simplifications (done on Mathematica 13.2.1.0 with FullSimplify) this is given by \eqref{eq:variance} of the main text.

\section{Case of different correlation time for $M_{ij}(t)$ and $h_i(t)$}

\begin{figure}[h]
    \centering
    \includegraphics[width=0.9\linewidth]{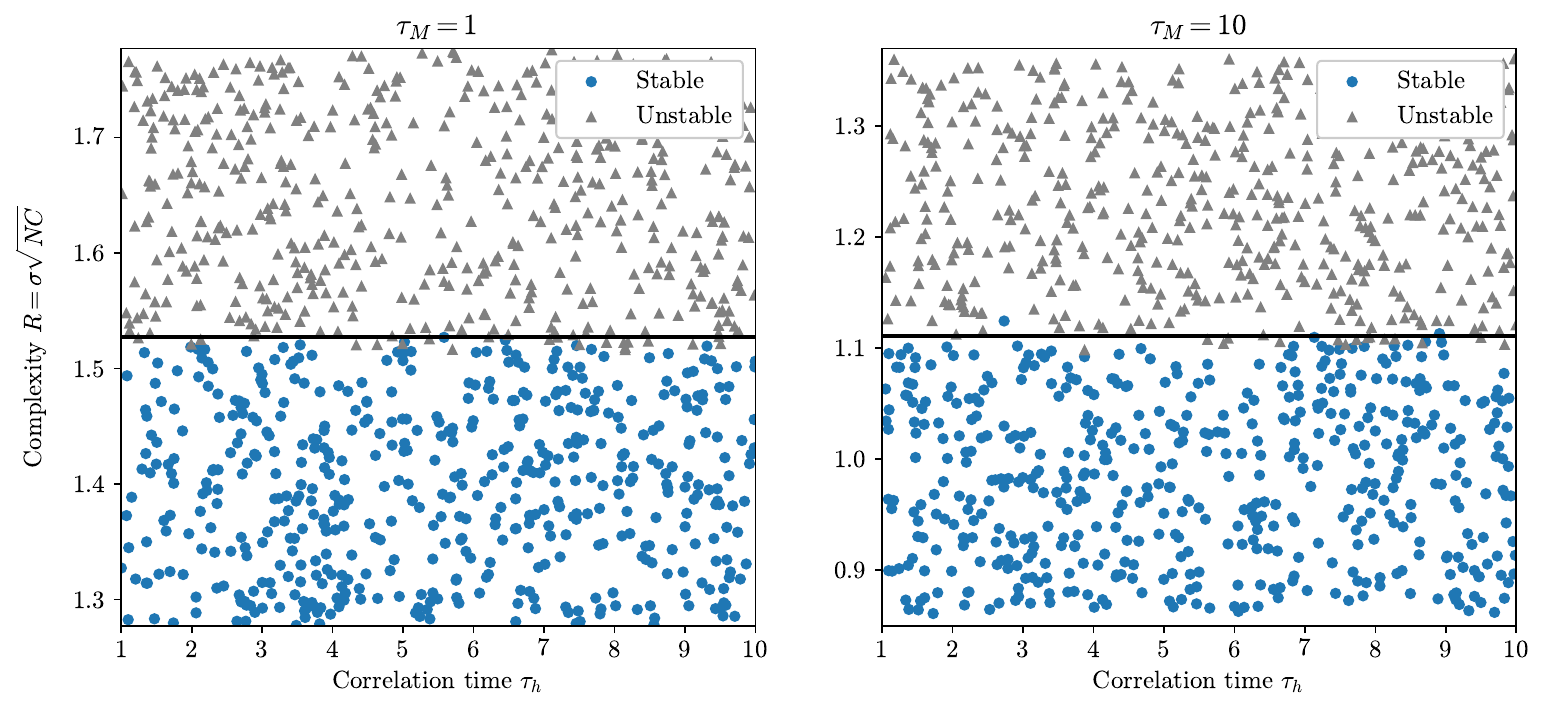}
    \caption{Numerical simulations of the linear model \eqref{eq:model} of the main text with parameters $M_{ij}(t)$ and $h_i(t)$ having different correlation times $\tau_M$ and $\tau_h$, as prescribed by \eqref{eq:different-Mh}. The left panel shows $\tau_M=1$, and the right panel shows $\tau_M=10$. The solid line is the theoretical result given by \eqref{eq:stability-bound} of the main text with $\tau=\tau_M$. Each simulation is classified as stable if it reaches a finite stationary state or unstable if it diverges. As expected, stability is independent of the parameter $\tau_h$, which results in a constant stability line as a function of this parameter.}
    \label{fig:tauM_tauh}
\end{figure}

In the main text we assumed for simplicity of presentation that $M_{ij}(t)$ and $h_i(t)$ were colored noises with the same correlation time. We relax here this assumption, by considering the case in which
\begin{equation}
\begin{aligned}
    \avg{M_{ij}(t)M_{ij}(t')} &= \sigma^2 Q(t-t'|\tau_M) \\
    \avg{h_i(t)h_i(t')} &= \rho^2 Q(t-t'|\tau_h)
\end{aligned}
\label{eq:different-Mh}
\end{equation}
where $Q(t|\tau) = e^{-\abs{t}/\tau}$. The procedure to find the stability criterion is analogous to the one presented in the main text. The corresponding DMFT equation remains the same as in the main text and is given by \eqref{eq:DMFT} of the main text, but with noise correlations that generalize to
\begin{equation}
\begin{aligned}
    \avg{\eta(t)\eta(t')} &= Q(t-t'|\tau_M) \avg{\delta x(t) \delta x(t')} \\
    \avg{h(t)h(t')} &= Q(t-t'|\tau_h)
\end{aligned}
\end{equation}
We can follow the same steps of the main text and find the following generalization of \eqref{eq:correlation-ODE} of the main text:
\begin{equation}
    -\ddot{C}\st(t) + \left[1 - R^2 Q(t|\tau_M)\right]C\st(t) = \rho^2 Q(t|\tau_h).
    \label{SMeq:correlation-ODE}
\end{equation}
Two independent solutions of the corresponding homogeneous ODE are given by \eqref{SMeq:u1-u2} with the identification $\tau=\tau_M$. Since the Wronskian of $u_1(t)$ and $u_2(t)$ is $1/\pi\tau_M$, by the method of variation of parameters the general solution of \eqref{SMeq:correlation-ODE} is
\begin{equation}
    C\st(t) = c_1u_1(t) + c_2u_2(t) + A_1(t)u_1(t) + A_2(t)u_2(t)
    \label{SMeq:Cst-solution}
\end{equation}
where
\begin{equation}
\begin{aligned}
    A_1 (t) 
    &= -\pi\tau_M \rho^2 \int dt e^{-t/\tau_h}u_2(t) \\
    &= \frac{\pi\tau_M \rho^2 (\tau_M R)^{-2\tau_M}}{(1-1/\tau_h)\sin(2\pi\tau_M)\Gamma(1-2\tau_M)} e^{(1-1/\tau_h)t} \hg{1}{2}(-(1-1/\tau_h)\tau_M;1-2\tau_M,1-(1-1/\tau_h)\tau_M;-\tau_M^2R^2 e^{-t/\tau_M}) \\
    &\qquad+ \frac{\pi\tau_M \rho^2 (\tau_MR)^{2\tau_M}}{(1+1/\tau_h)\tan(2\pi\tau_M)\Gamma(1+2\tau_M)}e^{-(1+1/\tau_h)t} \hg{1}{2}((1+1/\tau_h)\tau_M;1+2\tau_M,1+(1+1/\tau_h)\tau_M;-\tau_M^2R^2 e^{-t/\tau_M})
\end{aligned}
\end{equation}
and
\begin{equation}
\begin{aligned}
    A_2(t)
    &= \pi\tau_M\rho^2 \int dt e^{-t/\tau_h} u_1(t) \\
    &= -\frac{\pi\tau_M\rho^2(\tau_M R)^{2\tau_M}}{\Gamma(1+2\tau_M)(1+1/\tau_h)} e^{-(1+1/\tau_h)t} \hg{1}{2}((1+1/\tau_h)\tau_M;1+2\tau_M,1+(1+1/\tau_h)\tau_M;-\tau_M^2R^2e^{-t/\tau_M})
\end{aligned}
\end{equation}
These integrals were computed with the substitution $z = 2\tau R e^{-t/2\tau}$ and known integrals of Bessel functions. We want to impose the boundary conditions, which are $\dot{C}\st(0)=0$ and that $C\st(\infty)$ is finite. To do this we use the asymptotics of each term appearing in \eqref{SMeq:Cst-solution} which are
\begin{equation}
\begin{aligned}
    u_1(t) &\sim \frac{(\tau_M R)^{2\tau_M}}{\Gamma(1+2\tau_M)} e^{-t} \\
    u_2(t) &\sim - \frac{\Gamma(2\tau_M)(\tau_M R)^{-2\tau_M}}{\pi} e^t \\
    A_1(t)u_1(t) &\sim \frac{\rho^2}{2(1-1/\tau_h)} e^{-t/\tau_h} \\
    A_2(t)u_2(t) &\sim \frac{\rho^2}{2(1+1/\tau_h)} e^{-t/\tau_h}
\end{aligned}
\end{equation}
To have a finite $C\st(\infty)$ we impose that $c_2=0$. With this we also get that $C\st(\infty)=0$. Imposing $\dot{C}\st(0)=0$ fixes the value of $c_1$, which is not reported here. Setting $t=0$ in the resulting $C\st(t)$ one obtaines after simplifications (done on Mathematica 13.2.1.0 with FullSimplify) the following generalization of \eqref{eq:variance} of the main text:
\begin{equation}
    \sigma\st^2 = \frac{\rho^2}{1+1/\tau_h} \frac{\hg{1}{2}((1+1/\tau_h)\tau_M;1+2\tau_M,1+(1+1/\tau_h)\tau_M;-\tau_M^2R^2)}{2\hg{0}{1}(2\tau_M;-\tau_M^2R^2)-\hg{0}{1}(1+2\tau_M;-\tau_M^2R^2)}
\end{equation}
One can see that setting $\tau_h=\tau_M$ \eqref{eq:variance} of the main text is recovered after some manipulations. We also see that the denominator coincides with the one of \eqref{eq:variance} of the main text with the identification $\tau=\tau_M$. This implies, as mentioned in the main text, that stability depends on the correlation time of the interaction matrix only, and the stability criterion remains exactly the same as in the main text, which is given by \eqref{eq:stability-bound}. Figure~\ref{fig:tauM_tauh} shows numerical simulations that confirm this result.

\section{Case of correlated $M_{ij}(t)$ and $M_{ji}(t)$}

\begin{figure}[h]
    \centering
    \includegraphics[width=0.9\linewidth]{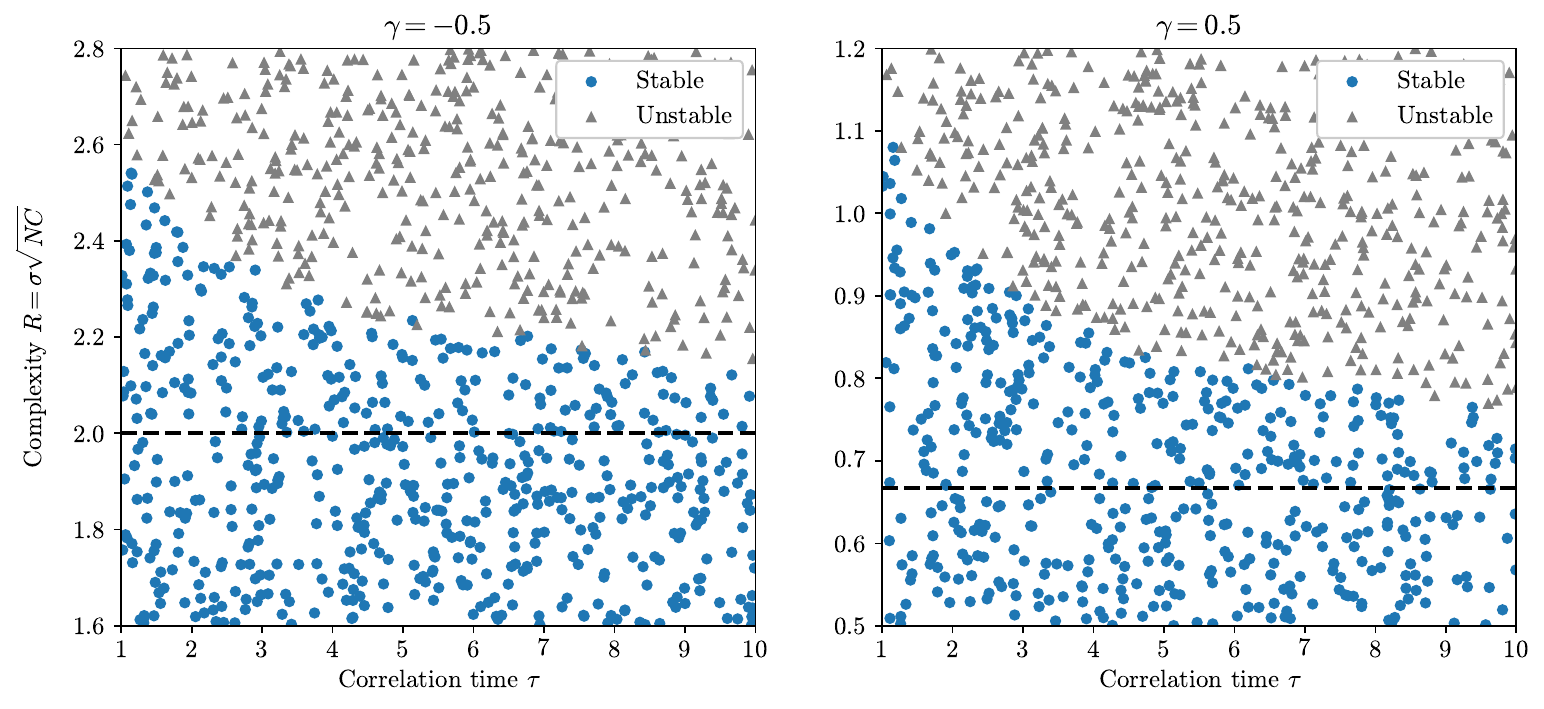}
    \caption{Numerical simulations of the linear model \eqref{eq:model} of the main text with an interaction matrix correlated as given by \eqref{eq:corr}. The left panel shows the case $\gamma=-0.5$, and the right panel shows $\gamma=0.5$. Each simulation is classified as stable if it reaches a finite stationary state or unstable if it diverges. The dashed line indicates the static bound $R=1/(1+\gamma)$. We observe that stability is consistently maintained beyond this bound, as in the $\gamma=0$ case.}
    \label{fig:gamma}
\end{figure}

In the main text, we assumed that each entry of the matrix $M_{ij}(t)$ evolves independently. We relax this assumption here by considering the case in which
\begin{equation}
    \textrm{corr}(M_{ij}(t), M_{ji}(t')) = \gamma Q(t-t')
\label{eq:corr}
\end{equation}
The elliptic law \cite{sommers1988spectrum} implies that at fixed time the spectrum of the matrix $M_{ij}(t)$ is uniformly supported in an ellipse centered at $(-1,0)$ whose real and imaginary axes are $(1+\gamma)R$ and $(1-\gamma)R$ respectively. In particular, the matrix $M_{ij}(t)$ has a fraction of eigenvalues with a positive real part whenever $R>1/(1+\gamma)$. The DMFT \eqref{eq:DMFT} of the main text could be generalized to the case $\gamma\neq0$, but we perform a purely numerical analysis. Operationally, we generate independent processes $z_{ij}(t)$ with correlations $\avg{z_{ij}(t) z_{ij}(t')} = Q(t-t')$ and obtain the matrix entries as
\begin{equation}
\begin{aligned}
    M_{ij}(t) &= \sigma z_{ij}(t) \\
    M_{ji}(t) &= \sigma\gamma z_{ij}(t) + \sigma \sqrt{1-\gamma^2} z_{ji}(t)
\end{aligned}
\end{equation}
We then integrate the linear model given by \eqref{eq:model} of the main text to determine if it converges to a finite stationary state or if it diverges. The results are shown in \figref{fig:gamma}. We observe the same phenomenology as the $\gamma=0$ model discussed in the main text. Stability is maintained beyond the classical limit $R=1/(1+\gamma)$ (dashed line) for all values of the correlation time of the disorder. Moreover, more rapidly fluctuating interactions lead to greater stability.

\end{document}